\documentclass[11pt]{article}

\usepackage[dvips]{graphicx} 
\usepackage[bookmarks=true]{hyperref}
\usepackage{epsfig,psfrag}
\hypersetup{breaklinks=true}

\let\fn\footnote
\renewcommand{\footnote}[1]{\linespread{1.1}\fn{#1}\linespread{1.29}}

\usepackage[left]{lineno}

\makeatletter\renewcommand{\section}{\@startsection
{section}{1}{\z@}{-3.5ex plus -1ex minus
    -.2ex}{2.3ex plus .2ex}{\bf }}
\makeatletter\renewcommand{\subsection}{\@startsection{subsection}{2}{\z@}{-3.25ex
plus -1ex minus
   -.2ex}{1.5ex plus .2ex}{\it }}
\makeatletter\renewcommand{\subsubsection}{\@startsection{subsubsection}{3}{-2.45ex}{-3.25ex
plus -1ex minus -.2ex}{1.5ex plus .2ex}{\it }}
\renewcommand{\thesection}{\arabic{section}.}
\renewcommand{\thesubsection}{\arabic{section}.\arabic{subsection}.}

\renewcommand{\theequation}{\thesection\arabic{equation}}
\makeatletter \@addtoreset{equation}{section}

\renewenvironment{thebibliography}[1]
     {\baselineskip=16pt plus 2pt minus 1pt
      \section*{\large\refname
        \@mkboth{\MakeUppercase\refname}{\MakeUppercase\refname}}%
     \list{\@biblabel{\@arabic\c@enumiv}}%
           {\settowidth\labelwidth{\@biblabel{#1}}%
            \leftmargin\labelwidth
            \advance\leftmargin\labelsep
            \@openbib@code
            \usecounter{enumiv}%
            \let\p@enumiv\@empty
            \renewcommand\theenumiv{\@arabic\c@enumiv}}%
      \sloppy
      \clubpenalty4000
      \@clubpenalty \clubpenalty
      \widowpenalty4000%
      \sfcode`\.\@m}

\newcommand{\acknowledgements}{\section*{Acknowledgements}
\addcontentsline{toc}{section}{\hspace{0.6cm}{\bf Acknowledgements}}}

\newcommand{\appendices}{\pdfbookmark[1]{Appendix}{appendices}
\section*{Appendix}\label{appendices}\setcounter{subsection}{0}
\setcounter{equation}{0}
\renewcommand*{\theHequation}{GG.\Roman{subsection}.\arabic{equation}}
\renewcommand{\thesubsection}{\Alph{subsection}.}
\makeatletter \@addtoreset{equation}{subsection}
\makeatother
\renewcommand{\theequation}{\thesubsection\arabic{equation}}
}

\usepackage{epsfig,rotating}
\usepackage{amsmath,amssymb}
\usepackage{amsfonts}
\usepackage{mathrsfs}
\usepackage{bbm}
\usepackage{bm}

\def\slasha#1{\setbox0=\hbox{$#1$}#1\hskip-\wd0\hbox to\wd0{\hss\sl/\/\hss}}

\def\periodb#1{\setbox0=\hbox{$#1$}#1\hskip-\wd0\hbox to\wd0{-}}

\newcommand{\unit}{\mathbbm{1}}   			

\newcommand{\CC}{\mathcal{C}}

\newcommand{\CCD}{\mathscr{D}}

\newcommand{\CH}{\mathcal{H}}
\newcommand{\CCH}{\mathscr{H}}
\newcommand{\CI}{\mathcal{I}}
\newcommand{\CJ}{\mathcal{J}}

\newcommand{\CL}{\mathcal{L}}

\newcommand{\CO}{\mathcal{O}}

\newcommand{\CT}{\mathcal{T}}

\newcommand{\CV}{\mathcal{V}}

\newcommand{\FR}{\mathbbm{R}}     			
\newcommand{\FC}{\mathbbm{C}}     			
\newcommand{\CPP}{{\mathbbm{C}P}}    			
\newcommand{\ah}{\hat{a}}

\newcommand{\dd}{\mathrm{d}}     			
\newcommand{\dpar}{\partial}     			
\newcommand{\de}{\mathrm{e}}     			
\newcommand{\di}{\mathrm{i}}     			

\newcommand{\bz}{{\bar{z}}}

\newcommand{\eand}{{~~~\mbox{and}~~~}}     		

\newcommand{\tr}{\,\mathrm{tr}\,}     			

\newcommand{\au}{\mathfrak{u}}

\newcommand{\sU}{\mathsf{U}}     			

\newcommand{\sSU}{\mathsf{SU}}

\newcommand{\sEnd}{\mathsf{End}\,}

\newcommand{\remark}[1]{}     				
\def\tyng(#1){\hbox{\tiny$\yng(#1)$}}			
\def\tyoung(#1){\hbox{\tiny$\young(#1)$}}			
\newcommand{\cpv}{\int\hspace{-0.4cm}-\,}

\begin{document}
\begin{titlepage}
\begin{flushright}
 HWM--10--36 \\ EMPG--10--26 \\ ZMP--HH/10--394
\end{flushright}
\vskip 2.0cm
\begin{center}
{\LARGE \bf Fuzzy Scalar Field Theory\\[0.4cm] as Matrix Quantum Mechanics}
\vskip 1.5cm
{\Large Matthias Ihl$^{\ast}$, Christoph Sachse$^{\dag}$ and Christian S{\"a}mann$^{\ddag}$}
\setcounter{footnote}{0}
\renewcommand{\thefootnote}{\arabic{thefootnote}}
\vskip 1cm
{\em $^{\ast}$ Instituto de F{\'i}sica\\
Universidade Federal do Rio de Janeiro\\
21941-972 Rio de Janeiro, RJ, Brasil}\\
{Email: {\ttfamily msihl@if.ufrj.br}}
\vskip 0.5cm
{\em $^{\dag}$ Fachbereich Mathematik\\
Bereich Algebra und Zahlentheorie\\
Universit{\"a}t Hamburg\\
D-20146 Hamburg, Deutschland}\\
{Email: {\ttfamily christoph.sachse@uni-hamburg.de}}
\vskip 0.5cm
{\em $^{\ddag}$ Department of Mathematics\\
Heriot-Watt University\\
Colin Maclaurin Building, Riccarton, Edinburgh EH14 4AS, U.K.\\
and Maxwell Institute for Mathematical Sciences, Edinburgh,
U.K.}\\
{Email: {\ttfamily C.Saemann@hw.ac.uk}}
\end{center}
\vskip 1.0cm
\begin{center}
{\bf Abstract}
\end{center}
\begin{quote}
We study the phase diagram of scalar field theory on a three dimensional Euclidean spacetime whose spatial component is a fuzzy sphere. The corresponding model is an ordinary one-dimensional matrix model deformed by terms involving fixed external matrices. These terms can be approximated by multitrace expressions using a group theoretical method developed recently. The resulting matrix model is accessible to the standard techniques of matrix quantum mechanics.
\end{quote}
\end{titlepage}

\section{Introduction and results}

Fuzzy spaces, such as the fuzzy sphere $S^2_F$ \cite{Berezin:1974du}, provide an interesting way of regularizing quantum field theories \cite{Madore:1991bw,Grosse:1995ar}: These spaces come with an algebra of functions which is finite dimensional. Correspondingly, the functional integral appearing in the partition function of a quantum field theory on a fuzzy space reduces to a finite dimensional integral. An advantage of this approach over a lattice regularization is that it preserves symmetries: the isometries of the classical manifold still have a well-defined action on the corresponding fuzzy space.

Interestingly, (scalar) quantum field theories on a fuzzy sphere\footnote{or, more generally, fuzzy projective algebraic varieties} correspond to hermitian matrix models with additional couplings to fixed external matrices \cite{Steinacker:2005wj,O'Connor:2007ea,Saemann:2010bw}. These matrices originate from the kinetic term in the action, and they yield an obstruction to applying the usual matrix model technology for reducing the partition function to an integral over eigenvalues. Nevertheless, one can use group theoretical methods to perform this reduction after Taylor-expanding the exponential of the kinetic term in the partition function \cite{O'Connor:2007ea,Saemann:2010bw}. The rewritten partition function can then be used to derive analytically the phase structure of scalar field theory on the fuzzy sphere and to compare the result to the numerical studies of \cite{Martin:2004un,GarciaFlores:2005xc,Panero:2006bx,Panero:2006cs,Das:2007gm,GarciaFlores:2009hf}: Qualitatively, the phase structures match and the position of a triple point appearing in the phase diagram roughly agreed in both the numerical and analytical results.

In this paper, we continue the work of \cite{O'Connor:2007ea,Saemann:2010bw} by studying scalar field theories on the three-dimensional spacetime $\FR\times S^2_F$. The resulting model is a special variant of matrix quantum mechanics with additional couplings to fixed external matrices. We can use again the group theoretical techniques developed in \cite{O'Connor:2007ea}. That is, we Taylor-expand the exponential of the spatial part of the kinetic term in the partition function and rewrite the resulting series order by order in terms of multitrace expressions. Re-exponentiating these yields the partition function of a matrix quantum mechanical model involving the multitrace terms. In fact, we can trivially translate the results obtained in \cite{Saemann:2010bw} to the matrix quantum mechanics model.

Similar to a pure hermitian matrix model, matrix quantum mechanics can be solved in the large $N$ limit. One approach is to map the system to a non-interacting Fermi gas \cite{Brezin:1977sv}, an alternative way is the collective field theory formalism developed in \cite{Jevicki:1979mb}. These techniques can still be applied if the potential contains multitrace terms of the matrix field, as done e.g.\ in \cite{Sugino:1994zr,Gubser:1994yb,Klebanov:1994pv} in the 2d gravity context. The model that we obtain will be more general in the deformations than the models described in these papers, but it will still be accessible to these techniques.

The purpose of this paper is to complement the numerical results on the phase diagram of scalar $\phi^4$-theory on $\FR\times S^2_F$ found in \cite{Bietenholz:2004xs,Medina:2005su,Medina:2007nv} by an analytical study. Using the techniques mentioned above, we are able to derive many properties of this phase diagram. In particular, we confirm the existence of a third phase as compared to pure matrix quantum mechanics. Furthermore, we find expressions for the various phase boundaries, and in an important region of the parameter space, we can write down explicit formulas for the free energy. In the remaining parameter space, we can calculate the free energy for each point using two simple numerical operations. 

This paper is structured as follows. In section 2, we briefly review scalar quantum field theories on fuzzy complex projective spaces and show how to turn these into matrix quantum mechanics. An expression for the free energy of such a theory on a fuzzy sphere is derived in section 3. In section 4, we discuss general features of the phase diagram of this theory and quantitative results are presented in section 5.

\section{Fuzzy scalar field theory as multitrace matrix quantum mechanics}	

\subsection{Fuzzy complex projective spaces}

We will work with Berezin-quantized complex projective space as described e.g.\ in \cite{Saemann:2010bw} and we briefly summarize the associated notions in the following. Consider the space $\FC^{n+1}$ with complex coordinates $z_\alpha$, $\alpha=0,...,n$. Using the fibration $\sU(1)\rightarrow S^{2n+1}\rightarrow \CPP^n$, we conclude that the set $\Sigma_\ell$ of functions on $\FC^{n+1}$ of the form
\begin{equation}\label{eq:functionInSigma}
 f(z)=\sum_{\alpha_i,\beta_i}f^{\alpha_1...\alpha_\ell\beta_1...\beta_\ell}\frac{z_{\alpha_1}...z_{\alpha_\ell}\bz_{\beta_1}...\bz_{\beta_\ell}}{|z|^{2\ell}}~,~~~f^{\alpha_1...\alpha_\ell\beta_1...\beta_\ell}\in\FC
\end{equation}
descends to a subset of the smooth functions on $\CPP^n\subset S^{2n+1}\subset \FC^{n+1}$. The Hilbert space $\CCH_\ell$ for Berezin-quantized $\CPP^n$ is the $\ell$-particle Hilbert space in the Fock space of $n+1$ harmonic oscillators. For functions in $\Sigma_\ell$ of the form \eqref{eq:functionInSigma}, the quantization prescription is explicitly given by 
\begin{equation}
 f(z)\mapsto \hat{f}=\sum_{\alpha_i,\beta_i}f^{\alpha_1...\alpha_\ell\beta_1...\beta_\ell}\frac{1}{\ell!}\ah^\dagger_{\alpha_1}...\ah^\dagger_{\alpha_\ell}|0\rangle\langle 0|\ah_{\beta_1}...\ah_{\beta_\ell}~.
\end{equation}
Note that real functions $f$ are mapped to hermitian operators $\hat{f}\in\sEnd(\CCH_\ell)$ and the constant function $f(z)=c$, $c\in\FR$, is mapped to $c\cdot\unit\in\sEnd(\CCH_\ell)$. Furthermore, $N_{n,\ell}:=\dim(\CCH_\ell)=\frac{(n+\ell)!}{n!\ell!}$ and a real function $f\in\Sigma_\ell$ can be interpreted as an $N_{n,\ell}\times N_{n,\ell}$-dimensional, hermitian matrix.
  
The Laplace operator $\Delta$ on $\Sigma$ can be lifted to $\sEnd(\CCH_\ell)$ and the lift is given by the quadratic Casimir $C_2$ of $\sSU(n+1)$ in the representation formed by $\sEnd(\CCH_\ell)$. One can show that \cite{IuliuLazaroiu:2008pk}
\begin{equation}
 \widehat{\Delta f}= C_2 \hat{f}~.
\end{equation}
Furthermore, we can define an integral operation on $\sEnd(\CCH_\ell)$ by taking the trace:
\begin{equation}
 \int \dd \mu_{FS}\, f=\frac{{\rm vol}(\CPP^n)}{N_{n,\ell}}\tr(\hat{f})~,
\end{equation}
where $\dd \mu_{FS}$ is the Liouville measure induced by the Fubini-Study metric on $\CPP^n$. For a more detailed exposition of these relations, see e.g.\ \cite{IuliuLazaroiu:2008pk}.

\subsection{Partition function of the model}

The action of scalar field theory with quartic potential on $\FR\times \CPP^n_F$ is given by
\begin{equation}\label{eq:baseaction}
 S[\Phi]=\beta\int \dd t\, \tr\left(\tfrac{1}{2}\Phi(t)\left(C_2-\dpar_t^2\right)\Phi(t)+r\,\Phi^2(t)+g\Phi^4(t)\right)~.
\end{equation}
The scalar fields are represented by the time-dependent, $N_{n,\ell}\times N_{n,\ell}$-dimensional hermitian matrices $\Phi(t)\in\CC^\infty(\FR,\sEnd(\CCH_\ell))$ and, as mentioned above, the spatial part of the Laplace operator $\Delta$ is given by the quadratic Casimir $C_2$ of $\sSU(n+1)$. Note that we are working with Euclidean time, which implies a different sign in front of the potential compared to Minkowski signature.

Explicitly, we have $C_2\Phi(t):=[L^i,[L^i,\Phi]]$, where the $L^i$ form generators of $\sSU(n+1)$ acting on $\CCH_\ell$. We follow the conventions of \cite{Saemann:2010bw} and normalize the generators according to 
\begin{equation*}
 [L_i,L_j]=:\di \sum_kf_{ijk}L_k~,~~\tr(L_i)=0~,~~\sum_iL_i^2=c_L \unit\eand\tr(L_iL_j)=\frac{c_L N_{n,\ell}}{(n+1)^2-1}\delta_{ij}~.  
\end{equation*}
This yields positive eigenvalues for $C_2$, and thus the action \eqref{eq:baseaction} is bounded from below for $g>0$.

The model \eqref{eq:baseaction} defines a one-dimensional field theory of matrix-valued scalar fields. Such theories are known as matrix quantum mechanics in the literature. The corresponding partition function reads
\begin{equation}
 Z=\int \CCD \Phi(t)\,\exp\left(-\beta S[\Phi]\right)~,
\end{equation}
where $\CCD \Phi(t)$ is the measure over functions on $\FR$ taking values in the Lie algebra of hermitian matrices of size $N_{n,\ell}\times N_{n,\ell}$. For each value of $t$, the measure corresponds to the Dyson measure\footnote{i.e.\ the standard translation-invariant measure induced by the bi-invariant Haar measure on $\sU(N_{n,\ell})$} on the space of hermitian matrices. As in the case of the pure matrix model, we can therefore split the partition function into an eigenvalue part $\CCD\Lambda(t)$ and an angular part $\CCD\Omega(t)$:
\begin{equation}
 Z=\int \CCD \Lambda(t)\CCD\Omega(t)\,\exp\left(-\beta S[\Phi]\right)~.
\end{equation}
One now usually exploits the fact that expressions in the action which consist exclusively of traces of polynomials in the $\Phi$ are independent of the angular part: Diagonalizing $\Phi$ according to $\Lambda:=\Omega^\dagger\,\Phi\Omega$, where $\Lambda$ is the diagonal matrix of eigenvalues of $\Phi$ and $\Omega$ is some unitary matrix, yields the simplification
\begin{equation}
 \tr(\Phi^n)=\tr((\Omega\Phi\Omega^\dagger)^n)=\tr(\Lambda^n)~.
\end{equation}
The fixed external matrices $L_i$ originating from the quadratic Casimir, however, present an obstacle to rewriting the action in terms of $\Lambda$. In \cite{O'Connor:2007ea} the same problem was analyzed for scalar field theory on fuzzy spaces, that is, the dimensional reduction of the model \eqref{eq:baseaction}. There it was suggested to perform a Taylor expansion of the exponential of the kinetic term in the partition function. Order by order, the terms in this expansion can be evaluated analytically using group theoretical methods and, when recombined, yield the partition function of a matrix model with an action containing multitrace terms. We can apply the same method here and directly take over the results obtained in \cite{Saemann:2010bw}: The Lagrangian of our model is given by 
\begin{equation}\label{eq:lag1}
\CL=\tr\Big(-\Phi\dpar_t^2\Phi+r\Phi^2+g\Phi^4\Big)+\CL_{\rm Casimir}~,
\end{equation}
where $\CL_{\rm Casimir}=\tr(\Phi(t)C_2\Phi(t))$. For the fuzzy sphere $\CPP^1_F$, the latter term can be approximated in the limit of large matrix sizes by
\begin{equation}\label{eq:rewriting}
\CL_{\rm Casimir}=\left(-\frac{1}{N}\tr(\Phi)-\frac{\beta}{3N^3}\tr(\Phi)^3\right)\tr(\Phi)+\left(1-\frac{\beta}{3N}\tr(\Phi^2)+\frac{2\beta}{N^2}\tr(\Phi)^2\right)\tr(\Phi^2)~,
\end{equation}
where we abbreviated $N=N_{1,\ell}$. We thus arrive at a standard one-dimensional matrix model, deformed by the multitrace terms contained in $\CL_{\rm Casimir}$. The corresponding actions for $\CPP^n_F$ with $n>1$ are easily deduced from the results in \cite{Saemann:2010bw}, too: They merely correspond to replacing the coefficients in \eqref{eq:rewriting} by more complicated expressions in $N$. Since our goal is primarily to reproduce the numerical results of \cite{Bietenholz:2004xs,Medina:2005su}, we will focus our attention on the case of the fuzzy sphere in the following.

\section{Evaluation of the free energy}

In this section, we use collective field theory \cite{Jevicki:1979mb} to compute the free energy of our one-dimensional matrix model. To leading order in the matrix size, this method is equivalent to the semi-classical approximation of the model as presented in \cite{Brezin:1977sv}, see also \cite{Ginsparg:1993is} for a nice review. 

\subsection{The free energy of multitrace matrix quantum mechanics}
\label{sect:multiqm}

We start from the general one-dimensional matrix model with Lagrangian
\begin{equation}\label{eq:ActionPureMM}
\begin{aligned}
 \CL&=\tfrac{1}{2}\tr(\dot{\Phi}^2)+u_2\tr(\Phi^2)+u_4\tr(\Phi^4)\\&\hspace{1.5cm}+v_{12}\tr(\Phi)^2+v_{14}\tr(\Phi)^4+v_{22}\tr(\Phi^2)^2+w_{12}\tr(\Phi)^2\tr(\Phi^2)\\
&=\tfrac{1}{2}\tr(\dot{\Phi}^2)+V~.
\end{aligned}
\end{equation}
For simplicity, we decompose the matrix $\Phi$ into $\Phi=\sum_a \Phi^a\tau_a$, where $\tau_a$ are hermitian generators of $\au(N)$, normalized according to $\tr(\tau_a\tau_b)=\delta_{ab}$. The Hamiltonian corresponding to \eqref{eq:ActionPureMM} can be written as
\begin{equation}
\begin{aligned}
 \CH=-\tfrac{1}{2}\sum_a\frac{\dpar^2}{\dpar \Phi^2_a}+&u_2\tr(\Phi^2)+u_4\tr(\Phi^4)\\+&v_{12}\tr(\Phi)^2+v_{14}\tr(\Phi)^4+v_{22}\tr(\Phi^2)^2+w_{12}\tr(\Phi)^2\tr(\Phi^2)~.
\end{aligned}
\end{equation}
We now switch to the collective fields $\phi(\lambda)\in \CC_c(\FR)$ via the Fourier transform
\begin{equation}\label{def:collectivefield}
 \phi(\lambda)=\int\frac{\dd k}{2\pi N}\, \de^{\di k \lambda}\tr(\de^{-\di k \Phi})~,
\end{equation}
where we inserted a factor of $\frac{1}{N}$ compared to \cite{Jevicki:1979mb} to facilitate the large $N$ limit. Here, the collective field $\phi(\lambda)$ will turn out to play a similar r{\^o}le as the eigenvalue density in the case of the hermitian matrix model. We correspondingly introduce the various moments of $\phi(\lambda)$:
\begin{equation}
 c_k:=\int \dd \lambda\,\phi(\lambda)\lambda^k~.
\end{equation}
Note that \eqref{def:collectivefield} implies that the collective field $\phi(\lambda)$ satisfies the normalization condition
\begin{equation}
 c_0=\int \dd \lambda\,\phi(\lambda)=\frac{1}{N}\tr(\unit)=1~.
\end{equation}
The potential $V$, i.e.\ the non-derivative terms in the Lagrangian \eqref{eq:ActionPureMM}, is easily seen to transform into 
\begin{equation}
V=\frac{1}{N}\int \dd \lambda\,\phi(\lambda)\Big((v_{12}c_1+v_{14}c_1^3)\lambda+(u_2+v_{22}c_2+w_{12}c_1^2)\lambda^2+u_4\lambda^4\Big) ~.
\end{equation}
The transformation of the kinetic term is technically more involved, see \cite{Jevicki:1979mb}. Here, let us just note that one eventually arrives at the transformed Hamiltonian $\CH=\CT+\CV$ where the effective potential $\CV$ reads as
\begin{equation}
 \CV=\int \dd \lambda\,\phi(\lambda)\big(\tfrac{1}{2N}G^2(\lambda,\phi)\big)+V 
\end{equation}
with the resolvent $G(\lambda,\phi)=\int \dd \zeta\,\frac{\phi(\zeta)}{\zeta-\lambda}$. To determine the collective field $\phi_0(\lambda)$ for the ground state of the system, we have to minimize the functional
\begin{equation}\label{eq:energy}
\begin{aligned}
 E(\mu_F,\nu_F,\kappa_F,\phi)=\CV&+\mu_F\left(1-\int \dd \lambda\,\phi(\lambda)\right)\\&+\nu_F\left(c_1-\int \dd \lambda\,\phi(\lambda)\lambda\right)+\kappa_F\left(c_2-\int \dd \lambda\,\phi(\lambda)\lambda^2\right)~.
\end{aligned}
\end{equation}
The variations with respect to $\phi$, $\mu_F$, $\nu_F$, $\kappa_F$, $c_1$ and $c_2$ yield the system of equations
\begin{equation}\label{eq:eoms}
 \begin{aligned}
  \tfrac{1}{2}&\left(\cpv\dd\zeta\,\frac{\phi(\zeta)}{\zeta-\lambda}\right)^2-\cpv\dd\zeta\,\frac{\phi(\zeta)}{\zeta-\lambda}\cpv\dd\xi\,\frac{\phi(\xi)}{\xi-\zeta}\\&\hspace{1cm}=\mu_F+\nu_F\lambda+\kappa_F\lambda^2-\Big((v_{12}c_1+v_{14}c_1^3)\lambda+(u_2+v_{22}c_2+w_{12}c_1^2)\lambda^2+u_4\lambda^4\Big)~,\\
  &\hspace{2.5cm}1=\int \dd \lambda\,\phi(\lambda)~,~~~
c_1=\int \dd \lambda\,\phi(\lambda)\lambda~,~~~
c_2=\int \dd \lambda\,\phi(\lambda)\lambda^2~,\\
&\hspace{3.5cm}\nu_F=-v_{12}c_1-3v_{14}c_1^3-2w_{12}c_1c_2~,~~~
  \kappa_F=-v_{22}c_2~,\hspace{2.0cm}
 \end{aligned}
\end{equation}
which is solved by standard methods, cf.\ \cite{Jevicki:1979mb,Shapiro:1980vx}. We arrive at the collective field
\begin{subequations}\label{eq:sol}
\begin{equation}
 \phi_0(\lambda)=\left\{\begin{array}{ll}\dfrac{1}{\pi}\sqrt{2\left(\mu_F-a_1\lambda-a_2\lambda^2-a_4\lambda^4\right)}&\mbox{for }\lambda\in\CI~, \\[0.2cm]
0 & \mbox{else}~,
\end{array}\right.
\end{equation}
where 
\begin{equation}
 a_1=2v_{12}c_1+4v_{14}c_1^3+2w_{12}c_1c_2~,~~~a_2=u_2+2v_{22}c_2+w_{12}c_1^2~,~~~a_4=u_4~,
\end{equation}
\end{subequations}
and $\CI\subset \FR$ can be any union of closed intervals such that $\phi_0(\lambda)=0$ at the boundaries. Equations \eqref{eq:eoms} imply that
\begin{equation}
 \int \dd\lambda\,\phi_0(\lambda)\,G^2(\lambda,\phi_0)=\tfrac{1}{3}\int \dd\lambda\,\phi_0(\lambda)\left(2 \left(\mu_F-a_1\lambda-a_2\lambda^2-a_4\lambda^4\right)\right)=\frac{\pi^2}{3}\int \dd\lambda\,\phi_0^3~,
\end{equation}
and the potential $V$ reads as
\begin{equation*}
V=\frac{1}{N}\int \dd \lambda\,\phi_0(\lambda)\left(-\frac{\pi^2}{2}\phi_0^2(\lambda)+\mu_F-c_1 v_{12}\lambda-c_2 v_{22}\lambda^2-3c_1^3v_{14}\lambda+2c_1c_2w_{12}\lambda\right)~.
\end{equation*}
Together with \eqref{eq:sol}, we arrive at the following expression for the energy \eqref{eq:energy} in the ground state:
\begin{equation}\label{eq:freeenergy}
\begin{aligned}
 NE(\mu_F,\nu_F,\kappa_F,\phi_0)=&\mu_F-v_{12}c_1^2-v_{22}c_2^2-3v_{14}c_1^4+2w_{12}c_1^2c_2
\\&-\tfrac{1}{3}\int \frac{\dd\lambda}{\pi}\,\left( 2 \left(\mu_F-a_1\lambda-a_2\lambda^2-a_4\lambda^4\right)\right)^{\frac{3}{2}}~.
\end{aligned}
\end{equation}
While all the integrals appearing above can be expressed in terms of elliptic functions as done in appendix A, it is not possible to use them to find analytic expressions for $\mu_F$ in general. 

\subsection{Comments on the semiclassical approximation}

The semi-classical approximation employed in \cite{Brezin:1977sv} leads to the same result as the collective field theory method. Here, one considers the quantum mechanical problem of finding the eigenvalues of the $N$-particle Hamiltonian corresponding to the action \eqref{eq:ActionPureMM}, where $\Phi=\Phi^\dagger\in{\rm Mat}_{\FC}(N)$. As in the case of the zero-dimensional matrix model, one can switch to an eigenvalue formulation of the model. The Vandermonde determinant arising from this can be absorbed in the quantum mechanical wavefunction, rendering it totally antisymmetric. The ground state energy of this system is then found by using the standard description in terms of free fermions, where the Lagrange multiplier $\mu_F$ of the collective field theory method becomes the Fermi energy. 

The only new ingredient here is the treatment of the multitrace terms: By linearizing e.g.\ $\tr(\Phi^2)$ around the vacuum expectation value $c_2=\langle \tr(\Phi^2)\rangle$, we find the relation
\begin{equation}
 (\tr(\Phi^2))^2\approx 2\langle \tr(\Phi^2)\rangle\tr(\Phi^2)-\langle \tr(\Phi^2)\rangle^2~,
\end{equation}
cf.\ \cite{Gubser:1994yb}. This induces a constant shift of the free energy proportional to $\langle \tr(\Phi^2)\rangle^2$ as well as a doubling of the na{\"i}ve contribution of the term corresponding to $\tr(\Phi^2)^2$. In \cite{Gubser:1994yb}, this was justified from physics principles. In the previous section, we saw that precisely the same modifications arise in our model from integrating out the Lagrange multiplier $\kappa_F$. 

\subsection[The large N limit]{The large $N$ limit}

As usual for matrix models\footnote{A similar problem is encountered in quantum field theory on the lattice.}, the large $N$ limit is not unique, but requires fixing of the scaling behavior of all coupling constants and the eigenvalues. While mathematically all scalings which leave the potential positive definite are equally valid, the resulting models are physically very different. One physical constraint one usually imposes is that in the limit $N\rightarrow \infty$, the coupling constants approach critical values.

Our model is supposed to provide an approximation for scalar field theory on $\FR^1\times S^2$, and we expect all terms of the potential as well as the kinetic term to survive in the large $N$ limit without becoming dominant. This implies that the leading order of all the contributions to the action should be of order $\CO(N^0)$, subleading contributions of order $\CO(N^{-1})$ provide corrections to the large $N$ limit. An overall scaling of the action is irrelevant, as we can define the free energy with a corresponding power of $N$.

To obtain a homogeneous scaling between the approximations due to the kinetic term and the potential, we can use the scaling found in \cite{Saemann:2010bw}:
\begin{equation}
 \beta\rightarrow N^{-\frac{1}{2}}\beta~,~~~\lambda\rightarrow N^{-\frac{1}{4}}\lambda~,~~~r\rightarrow N^{2}r~,~~~g\rightarrow N^{\frac{5}{2}}g~,
\end{equation}
where $r$ and $g$ are the actual parameters of our model \eqref{eq:lag1}. In addition, we have to scale $t\rightarrow \frac{1}{N}t$ to include the right scaling of the temporal part of the kinetic term. Altogether, our model corresponds to the general model \eqref{eq:ActionPureMM} with the following parameters:
\begin{equation}
u_2=1+r~,~~~u_4=g~,~~~v_{12}=-1~,~~~v_{14}=-\frac{\beta}{3}~,~~~v_{22}=-\frac{\beta}{3}~,~~~w_{12}=\frac{2}{3}\beta~,
\end{equation}
and for $\lambda\in\CI$, the corresponding collective field reads as
\begin{equation*}
 \phi_0(\lambda)=\dfrac{1}{\pi}\sqrt{2\left(\mu_F-\left(-2c_1-\frac{4}{3}\beta c_1^3+\frac{4}{3}\beta c_1c_2\right)\lambda-\left(1+r-\frac{2}{3}\beta c_2+\frac{2}{3}\beta c_1^2\right)\lambda^2-g\lambda^4\right)}~.
\end{equation*}
We furthermore have the relation 
\begin{equation}
 a_1=2c_1(r-a_2)~.
\end{equation}

\section{The phase diagram}

\subsection{The three phases}

We will exclusively deal with the situation in which $r_0$ is negative enough for the potential to have a local maximum. Note that in matrix form, our potential is symmetric under $\Phi\rightarrow -\Phi$. In the collective field reformulation, we have a symmetry of the potential $V$ under $\lambda\rightarrow -\lambda$. Naively, one therefore expects that symmetric phases are the dominant ones. However, our experience with the pure matrix model \cite{Saemann:2010bw} suggests that we should also allow for a third phase. This is further motivated by the numerical findings of \cite{Bietenholz:2004xs,Medina:2005su,Medina:2007nv}, which also observe (at least) three different phases. Depending on $\mu_F$, we can distinguish three phases:
\begin{itemize}
 \item[I.] The {\em disordered phase} or {\em single-cut case}. Here, $\mu_F>0$ and the filling of the eigenvalues covers the local maximum completely. There is a single interval $\CI=[-\lambda_1,\lambda_1]$ over which one has to integrate $\lambda$, and it is symmetric around the origin. This implies that the first moment of the collective field vanishes: $c_1=0$.
 \item[II.] The {\em non-uniform ordered phase} or {\em symmetric double-cut case}. In this case, $\mu_F<0$ and the Fermi sea of eigenvalues splits up into two symmetric, disjoint pieces. That is, there are two intervals $\CI=[-\lambda_2,-\lambda_1]\cup[\lambda_1,\lambda_2]$ with $0\leq\lambda_1\leq\lambda_2$ as support for the integral, which are again symmetric around the origin. We again have $c_1=0$.
 \item[III.] The {\em uniform (ordered) phase} or {\em asymmetric double-cut case}. As in phase II, $\mu_F<0$ and the Fermi sea is split, but this time into two asymmetric pieces. The length of the two intervals $\CI_1\cup\CI_2=\CI$ is different. Note that true uniform ordering is only achieved in the totally asymmetric case, in which one of the intervals has shrunk to zero size. Here, $c_1\neq 0$.
\end{itemize}
The actual phase is determined from existence conditions and the fact that the physical system adopts the ground state with the lowest free energy. 

\begin{figure}[h]
\center
\begin{picture}(420,100)
\put(52.0,80.0){\makebox(0,0)[c]{$V(\lambda)$}}
\put(130.0,27.0){\makebox(0,0)[c]{$\lambda$}}
\put(197.0,80.0){\makebox(0,0)[c]{$V(\lambda)$}}
\put(275.0,27.0){\makebox(0,0)[c]{$\lambda$}}
\put(344.0,80.0){\makebox(0,0)[c]{$V(\lambda)$}}
\put(420.0,27.0){\makebox(0,0)[c]{$\lambda$}}
\includegraphics[width=47mm]{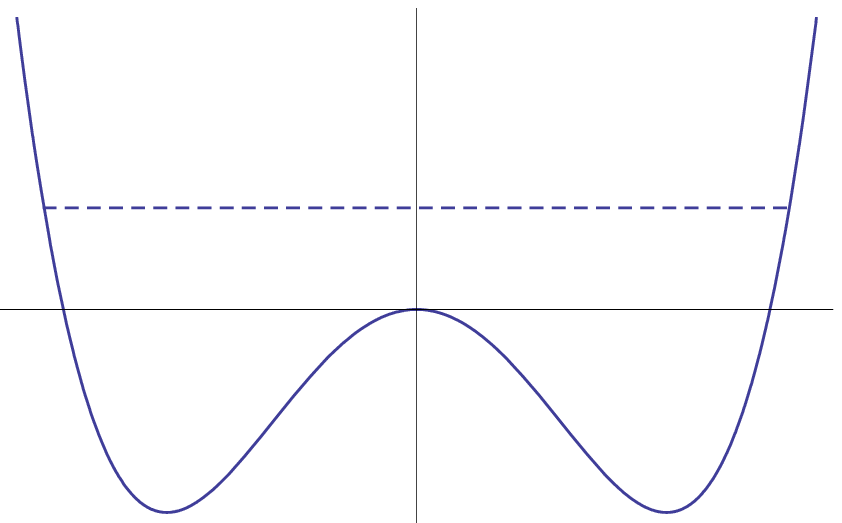}~~
\includegraphics[width=47mm]{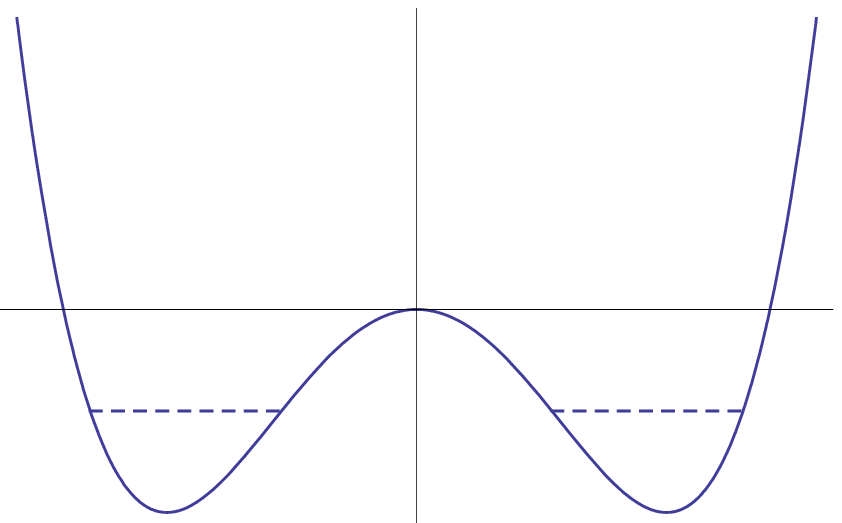}~~
\includegraphics[width=47mm]{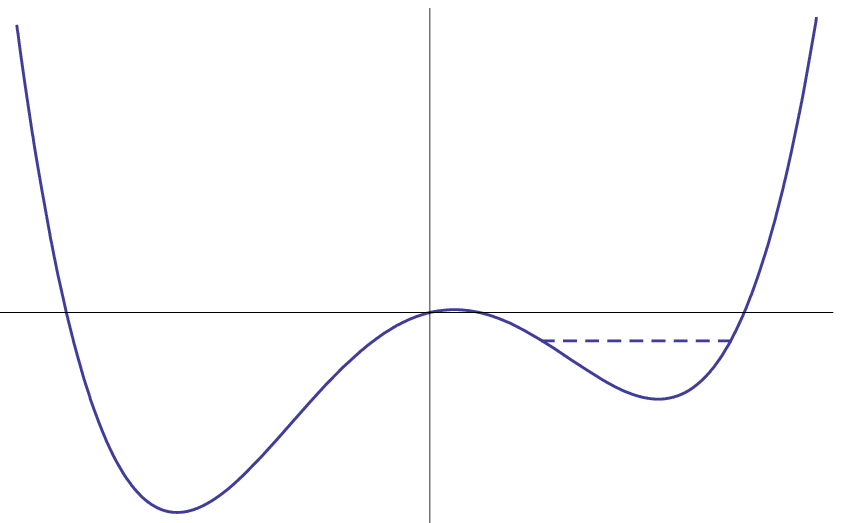}
\end{picture}
\caption{The three phases. The solid line describes the potential $V(\lambda)$ felt by the eigenvalues. The dashed lines mark the Fermi energy $\mu_F$. In phase III, we chose a totally asymmetric filling with $\CI_1=\emptyset$. The asymmetry of the potential in this phase is due to $c_1\neq 0$. }\label{fig:phases}
\end{figure}

Note that our model can always be treated as ordinary matrix quantum mechanics, where the coefficients of the potential are dependent on the moments $c_1$ and $c_2$. That is, to study our model, we can solve matrix quantum mechanics for a general potential and then derive self-consistency conditions on the moments. For this reason, it is possible to evaluate the exact location of the phase transition between phases I and II as well as the existence domain of phase III in our model analytically.

\subsection{The phase transition I to II}

The phase transition between the single-cut phase and the double-cut phase obviously occurs at $\mu_F=0$. In this case, the elliptic integrals can be performed explicitly. The interval $\CI$ on which $\phi_0(\lambda)$ is supported is given by
\begin{equation}
 \CI=\left(-\sqrt{\tfrac{-u_2-2v_{22}c_2}{u_4}},\sqrt{\tfrac{-u_2-2v_{22}c_2}{u_4}}\right)~,
\end{equation}
and the normalization condition yields
\begin{equation}\label{eq:normalization}
 \frac{\sqrt{8}\left(-u_2-2 v_{22}c_2\right)^{\frac{3}{2}}}{3\pi u_4}=1~.
\end{equation}
The second moment $c_2$ is determined by the condition
\begin{equation}
 c_2=\frac{\sqrt{32}\left(-u_2-2v_{22}c_2\right)^{\frac{5}{2}}}{15\pi u_4^2}=
\frac{(3\pi)^{\frac{2}{3}}}{5u_4^{\frac{1}{3}}}~,
\end{equation}
where we made use of \eqref{eq:normalization}. Plugging this into \eqref{eq:normalization} yields a phase transition at
\begin{equation}
 u_2=-\frac{(3\pi)^{\frac{2}{3}}\left(5u_4+4v_{22}\right)}{10 (u_4)^{\frac{1}{3}}}~.
\end{equation}
If $u_2$ is larger than the right-hand side, we are in the single-cut phase, while for $u_2$ smaller than the right-hand side, the system is in the double-cut phase. 

\subsection{Existence condition for phase II}\label{sec:ExPhaseII}

Similarly to the pure matrix model discussed in \cite{Saemann:2010bw}, there is a region of the parameter space which is not covered by phase I and where the self-consistency relation for $c_2$ cannot hold true in phase II. To see this, we compute $c_2$ for the double cut solution of pure matrix quantum mechanics with $v_{12}=v_{14}=v_{22}=w_{12}=0$ for arbitrary parameters $a_2<0$ and $a_4>0$. We then translate to our model by identifying 
\begin{equation}\label{eq:relationru2}
 a_2=1+r-\frac{2}{3}\beta c_2\big(a_2,a_4,\mu_F(a_2,a_4)\big)\eand a_4=u_4=g~,
\end{equation}
where $r$ and $g$ are the actual parameters of our model. The region of the parameter space in which we expect the self-consistency relation to be problematic corresponds to $0<a_4/(-a_2)\ll 1$. In this region, the wells in the potential are sufficiently deep to be approximated by a parabola. That is, we approximate e.g.\ the right well as follows
\begin{equation}
 2(\mu_F-a_2\lambda^2-a_4 \lambda^4)\approx 2\mu_F+\frac{a_2^2}{2g}+4 a_2\left(\lambda-\sqrt{\frac{-a_2}{2g}}\right)^2~.
\end{equation}
Both the integrals yielding the normalization condition for $\phi_0$ and the second moment $c_2$ can be computed and read as
\begin{equation}
\begin{aligned}
 \int \dd \lambda\, \phi_0(\lambda)&\approx \frac{a_2^2+4 g \mu_F}{4 g \sqrt{-a_2}}~,\\
 \int \dd \lambda\, \phi_0(\lambda)\lambda^2&\approx\frac{(a_2^2+4 g \mu_F)(17 a_2^2+4 g \mu_F)}{128 g^2 (-a_2)^{3/2}}~.
\end{aligned}
\end{equation}
We can use these results together with \eqref{eq:relationru2} to determine the function $r(a_2)$:
\begin{equation}
 r\approx-1+\frac{\beta}{12\sqrt{-a_2}}-\frac{1}{3}a_2\left(\frac{\beta}{g}-3\right)~.
\end{equation}
Note that the larger $|a_2|$, the better this approximation becomes. In particular, we see that for $g\leq\frac{1}{3}\beta$, $r$ grows as $a_2$ becomes more negative. This implies that phase II exists for arbitrarily large values of $|r|$ only if $g>\frac{1}{3}\beta$.

\subsection{Existence condition for a totally asymmetric phase III}

To simplify our analysis, we will identify the third phase with a totally asymmetric filling. That is, $\CI$ is again a single interval, filling only one of the two wells in the potential. It is obvious that the transition from phase I to the totally asymmetric filling has to go smoothly through all possible asymmetric fillings, as close to the boundary between phases I and II, no totally asymmetric solution will exist. 

The existence boundary for the totally asymmetric solution is another line in our phase diagram which can be determined analytically if we neglect the contribution of the odd moment $c_1$. As we will see later in our numerical studies, the odd moments decrease the depth of the filled well. The bound obtained by this approximation for the existence of a totally asymmetric phase III is therefore an upper bound.

In our approximation, this line is determined by the fact that one of the two wells of the potential is filled up to the local maximum and that a further increase in Fermi energy $\mu_F$ would lead to a spilling of eigenvalues into the other well. Consequently, we put $\mu_F=0$ and consider the interval
\begin{equation}
 \CI=\left(0,\sqrt{\tfrac{-u_2-2v_{22}c_2}{u_4}}\right)~.
\end{equation}
The computation of the existence boundary then proceeds exactly as the computation in the previous section. We obtain
\begin{equation}
  \frac{\sqrt{2}(-u_2-2v_{22}c_2 )^{\frac{3}{2}}}{3\pi u_4}=1\eand c_2=\frac{\sqrt{8}\left(-u_2-2v_{22}c_2\right)^{\frac{5}{2}}}{15\pi u_4^2}=
\frac{(6\pi)^{\frac{2}{3}}}{5u_4^{\frac{1}{3}}}~,
\end{equation}
and the existence domain of the totally asymmetric phase is given by
\begin{equation}
 u_2\leq-\frac{(3\pi)^{\frac{2}{3}}\left(5u_4+4v_{22}\right)}{5 (2u_4)^{\frac{1}{3}}}~.
\end{equation}

\subsection{Pure matrix quantum mechanics}

Before discussing the phase diagram of our model, let us briefly discuss the slightly simpler case of pure matrix quantum mechanics, for which $v_{12}=v_{14}=v_{22}=w_{12}=0$. Here, the phase transition between I and II occurs at
\begin{equation}
 u_4=-\frac{\sqrt{8}(-u_2)^{\frac{3}{2}}}{3\pi}~,
\end{equation}
and the existence boundary for the totally asymmetric cut is
\begin{equation}
 u_4=-\frac{\sqrt{2}(-u_2)^{\frac{3}{2}}}{3\pi}~.
\end{equation}
\begin{figure}[h]
\hspace{3.5cm}
\begin{picture}(240,185)
\psfrag{3.0}{\kern-3pt 3.0}
\psfrag{2.5}{\kern-3pt 2.5}
\psfrag{2.0}{\kern-3pt 2.0}
\psfrag{1.5}{\kern-3pt 1.5}
\psfrag{1.0}{\kern-3pt 1.0}
\psfrag{0.5}{\kern-3pt 0.5}
\psfrag{-}{-}
\psfrag{0}{0}
\psfrag{5}{\kern-2pt 5}
\psfrag{4}{\kern-2pt 4}
\psfrag{3}{\kern-2pt 3}
\psfrag{2}{\kern-2pt 2}
\psfrag{1}{\kern-2pt 1}
\includegraphics[scale=1.15]{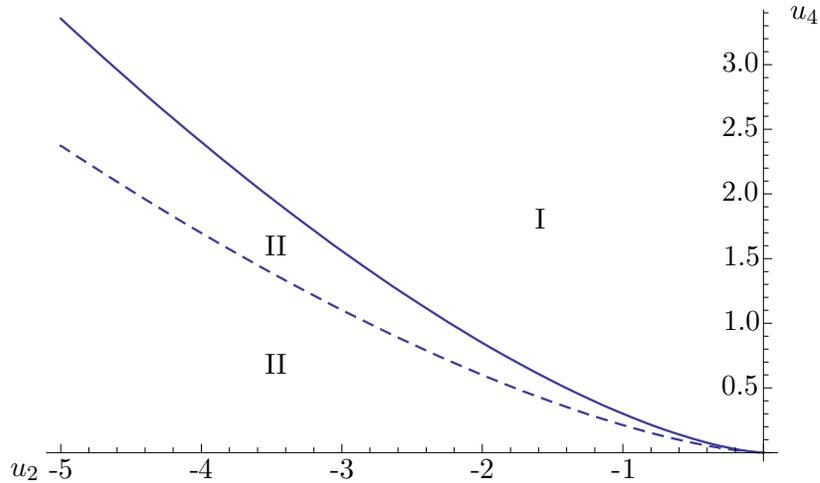}
\put(10.0,177.0){\makebox(0,0)[c]{$u_4$}}
\put(-285.0,4.0){\makebox(0,0)[c]{$u_2$}}
\put(-90.0,100.0){\makebox(0,0)[c]{I}}
\put(-190.0,90.0){\makebox(0,0)[c]{II}}
\put(-190.0,45.0){\makebox(0,0)[c]{II}}
\end{picture}
\vspace*{-5pt}
\caption{The phase diagram for pure matrix quantum mechanics. The solid line describes the phase transition and the dashed line the boundary of the existence domain of the totally asymmetric cut. The roman numerals describe the actual phases of the system.}\label{fig:phaseMQM}
\end{figure}

One can now show by a simple physical argument that the lowest energy configuration is always the symmetric double-cut solution, and thus phase III does not exist in pure matrix quantum mechanics: In the time-independent ground state and for large $N$, one would expect that moving one eigenvalue from one well to another does not cost or yield energy. This automatically implies that the Fermi energy in both wells is the same. To be rigorous, one can introduce filling fractions $\rho_1$ and $\rho_2$ for the left and right wells of the potential with $\rho_1+\rho_2=1$ and $\int_{\CI_i} \phi_0(\lambda)=\rho_i$. One can then determine the true minimum of the free energy numerically, which confirms the physical argument. The phase diagram of pure matrix quantum mechanics is depicted in figure \ref{fig:phaseMQM}.

\section{Results}

In the following, we put $\beta=\frac{1}{2}$ and perform an analysis of the phase diagram relying on solving the system of equations \eqref{eq:sol} and determining the free energy. As we already computed the location of the phase transition between phases I and II, it remains to compare the free energy on the overlap of the existence domains of phases II and III.

\subsection{Phases I and II}

Recall that the filling of the wells with eigenvalues is symmetric and therefore the odd moments of the collective field, and in particular $c_1$, vanish. Equations \eqref{eq:sol} cannot be solved analytically, as they involve elliptic integrals. We therefore apply the following algorithm:
\begin{enumerate}
\setlength{\itemsep}{-1mm}
 \item pick a value for $a_4=g$ and $a_2$
 \item evaluate $\mu_F$ for this pair by numerically minimizing $|1-\int_\CI\dd\lambda\,\phi_0(\lambda)|$
 \item numerically evaluate the second moment $c_2$
 \item deduce the value of $r$ from $a_2$ and $c_2$
 \item evaluate the free energy at the point $(r,g)$ in the parameter space
\end{enumerate}
By applying this algorithm to a range of values for $a_2$ and $a_4$, we can perform a sweep of the $r$-$g$-parameter space. The only restriction here is the existence boundary for phase II for $g\leq \frac{1}{3}\beta$.

The region of the $r$-$g$-parameter space, in which we expect an overlap between phases II and III is characterized by large $|r|$ and small $g$. The potential in this region has broad and deep wells, and correspondingly $\mu_F$ is small compared to the depth of the wells. This implies that an approximation of the potential by a parabola, like the one used in section \ref{sec:ExPhaseII}, should work reasonably well. 

\begin{figure}[h]
\hspace{2cm}\begin{picture}(320,200)
\psfrag{200}{\kern-2pt 200}
\psfrag{400}{\kern-2pt 400}
\psfrag{10}{\kern-4pt 10}
\psfrag{20}{\kern-4pt 20}
\psfrag{1.5}{\kern-2pt $g$}
\psfrag{1.0}{\kern-2pt 1.0}
\psfrag{0.5}{\kern-2pt 0.5}
\psfrag{-}{-}
\psfrag{0}{\kern-2pt 0}
\includegraphics[scale=0.7]{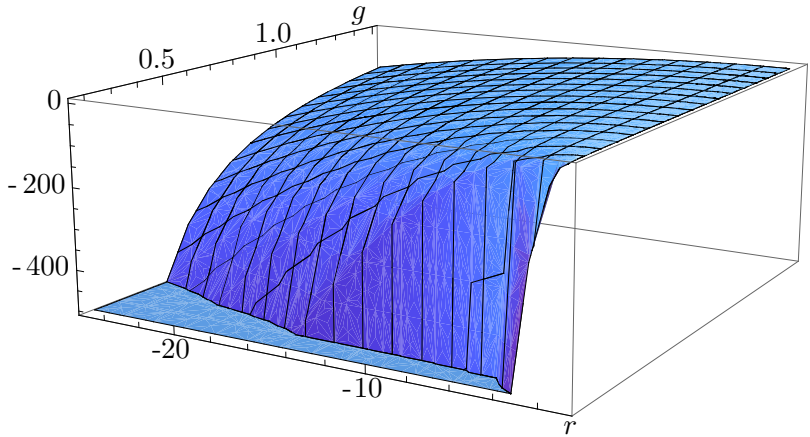}
\put(-90.0,30.0){\makebox(0,0)[c]{$r$}}
\end{picture}
\vspace*{-1.0cm}
\caption{The free energy in phase II as obtained from the approximation of the potential wells by parabolas.}\label{fig:FEsym}
\end{figure}

Explicitly, we approximate the collective field by
\begin{equation}\label{eq:approx}
 \phi_0(\lambda)\approx\left\{\begin{array}{ll}\dfrac{1}{\pi}\sqrt{p_0-\alpha(\lambda-\lambda_{\rm min})^2}&\mbox{for }\lambda\in\CI~, \\[0.2cm]
0 & \mbox{else}~.
\end{array}\right.
\end{equation}
The zeros of this collective field $\phi_0(\lambda)$ defining its support $\CI=[-\lambda_R,-\lambda_L]\cup[\lambda_L,\lambda_R]$ are
\begin{equation}
 \lambda_L=\lambda_{\rm min}-\sqrt{\frac{p_0}{\alpha}}\eand\lambda_R=\lambda_{\rm min}+\sqrt{\frac{p_0}{\alpha}}~.
\end{equation}
Together with the integrals given in appendix B, we can now compute 
\begin{equation}
 p_0=\sqrt{\alpha}~,~~~c_2=\frac{1}{4\sqrt{\alpha}}+\lambda_{\rm min}^2~,~~~a_2=1+r-\frac{\beta}{6\sqrt{\alpha}}-\frac{2}{3}\beta\lambda_{\rm min}^2~.
\end{equation}
From these relations, we derive
\begin{equation}
 \begin{aligned}
  r&=-1-\frac{\alpha}{4}+\frac{\beta}{6\sqrt{\alpha}}+\frac{2\beta\lambda_{\rm min}^2}{3}~,~~~g=\frac{\alpha}{8\lambda_{\rm min}^2}~,\\
  NE&=\frac{\sqrt{\alpha}}{4}+\frac{\beta}{48\alpha}-\frac{\alpha\lambda_{\rm min}^2}{8}+\frac{\beta\lambda_{\rm min}^2}{6\sqrt{\alpha}}+\frac{\beta\lambda_{\rm min}^4}{3}~.
 \end{aligned}
\end{equation}
Rewriting the free energy $NE$ in terms of $r$ and $g$ yields a lengthy but analytic expression, which is plotted in figure \ref{fig:FEsym}. Note that one can easily check the accuracy of the approximation by verifying the exact normalization condition as well as the self-consistency condition for $c_2$ using the approximate values of $\mu_F$ and $c_2$. This plot is confirmed by the numerical results of the first algorithm.

\subsection{Phase III}

Phase III is more intricate to deal with due to the additional appearance of the terms involving the odd moment $c_1$. From the expression of the free energy \eqref{eq:freeenergy}, it is clear that the free energy in phase III is much larger than the free energy in phase II if $c_2$ and the value of the integral are assumed to be roughly the same. We therefore do not expect phase III to compete with phase II, but to exist only where phase II does not. This is precisely the case in the region of the $r$-$g$-parameter space for which the approximation of the wells of the potential by parabolas, and thus \eqref{eq:approx}, works well.

\begin{figure}[h]
\hspace{2.4cm}\begin{picture}(320,180)
\psfrag{200}{200}
\psfrag{400}{400}
\psfrag{600}{600}
\psfrag{10}{\kern-4pt 10}
\psfrag{20}{\kern-4pt 20}
\psfrag{1.5}{\kern-2pt 1.5}
\psfrag{1.0}{\kern-2pt 1.0}
\psfrag{0.5}{\kern-2pt 0.5}
\psfrag{-}{-}
\psfrag{0}{0}
\includegraphics[scale=0.7]{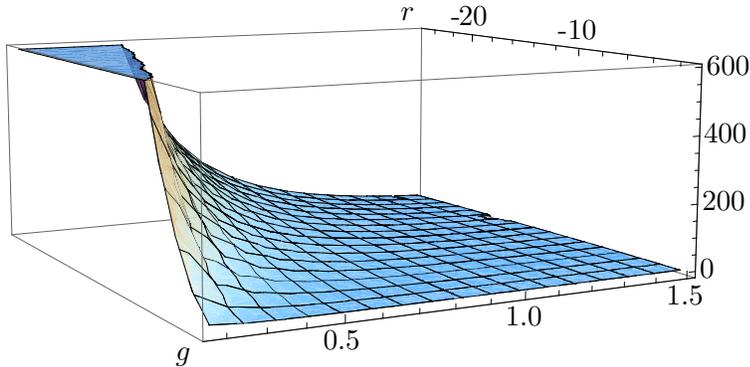}
\put(-125.0,165.0){\makebox(0,0)[c]{$r$}}
\put(-210.0,35.0){\makebox(0,0)[c]{$g$}}
\end{picture}
\vspace*{-1.3cm}
\caption{The free energy in phase III as obtained from the approximation of the potential wells by parabolas.}\label{fig:FEasym}
\end{figure}

For a totally asymmetric filling, i.e.\ a collective field $\phi_0(\lambda)$ with support $\CI=[\lambda_L,\lambda_R]$ we obtain
\begin{equation}
 p_0=2\sqrt{\alpha}~,~~~c_1=\lambda_{\rm min}~,~~~c_2=\frac{1}{2\sqrt{\alpha}}+\lambda_{\rm min}^2~,~~~a_2=1+r-\frac{\beta}{3\sqrt{\alpha}}~.
\end{equation}
This leads to the following expressions:
\begin{equation}
 \begin{aligned}
  r&=\frac{1}{12}\left(6-3\alpha-\frac{2\beta}{\sqrt{\alpha}}\right)~,~~~g=\frac{\alpha-2}{8\lambda_{\rm min}^2}+\frac{\beta}{12\sqrt{\alpha}\lambda_{\rm min}^2}~,\\
  NE&=\frac{12 \alpha^{3/2}+2\beta+6\alpha\lambda_{\rm min}^2-3\alpha^2\lambda_{\rm min}^2+30\sqrt{\alpha}\beta\lambda_{\rm min}^2+64\alpha\beta\lambda_{\rm min}^4}{24\alpha}~.
 \end{aligned}
\end{equation}
A plot of the free energy in phase III according to this approximation is given in figure \ref{fig:FEasym}.

\subsection{Discussion}

In the region of the parameter space in which phases II and III coexist, the free energy is negative in phase II, while it is positive in phase III. Note that this observation is due to our restriction of phase III to a totally antisymmetric eigenvalue filling. Actually, one would expect the system to adopt intermediate phases so that the free energy changes continuously between the extrema of a symmetric and a totally asymmetric filling. 

From the different signs of the free energy in phases II and III it follows that phase III is only adopted by the system if phase II is not available. For large $|r|$, we therefore expect a phase transition at the line $g=\tfrac{1}{3}\beta$, below which phase II does not necessarily exist. Interestingly, the same linear phase boundary  between phases II and III also appeared in the numerical results of \cite{Bietenholz:2004xs}. The fact that our phase boundary only holds for large $|r|$ is due to our approximation of the exponential of the kinetic term of the original model to second order. We expect higher orders to provide the necessary further corrections. Also, higher order contributions will yield terms in the potential of the one-dimensional matrix model which are of order $\tr(\Phi^n)$ with $n>4$. These terms would give rise to more than two wells in the potential leading to additional phases.

In summary, we conclude that the phase diagram of our theory is a slight deformation of that of matrix quantum mechanics, with the addition of a third phase. The phase boundaries between phases I and II and phases II and III are given, respectively, by the two curves
\begin{equation}
 r=-\frac{(3\pi)^{\frac{2}{3}}\left(5g-4\right)}{10 g^{\frac{1}{3}}}\eand g=\frac{1}{3}\beta~.
\end{equation}

\acknowledgements
                                                                                                                                  
C.\ S\"amann would like to thank Denjoe O'Connor for discussions. The work of M.\ Ihl was supported by a grant from CNPq (Brazilian funding agency). C.\ Sachse was partially supported by a postdoctoral research fellowship of the Deutsche Forschungsgemeinschaft while this work was completed. The work of C.\ S\"amann was supported by a Career Acceleration Fellowship from the UK Engineering and Physical Sciences Research Council. 

\appendices

\subsection{The elliptic integrals}

Throughout the paper, we encountered elliptic integrals, which can be reduced to the following ones:
\begin{equation}
\begin{aligned}
 \CI_{1}&=\int_a^b \dd \lambda\, \sqrt{(b^2-\lambda^2)(\lambda^2-a^2)}~,&\CJ_{1}&=\int_0^b \dd \lambda\, \sqrt{(b^2-\lambda^2)(\lambda^2+a^2)}~,\\
 \CI_{2}&=\int_a^b \dd \lambda\, \lambda^2\sqrt{(b^2-\lambda^2)(\lambda^2-a^2)}~,&\CJ_{2}&=\int_0^b \dd \lambda\, \lambda^2\sqrt{(b^2-\lambda^2)(\lambda^2+a^2)}~,\\
 \CI_{3}&=\int_a^b \dd \lambda\, \left(\sqrt{(b^2-\lambda^2)(\lambda^2-a^2)}\right)^3~,&\CJ_{3}&=\int_0^b \dd \lambda\, \left(\sqrt{(b^2-\lambda^2)(\lambda^2+a^2)}\right)^3~,
\end{aligned}
\end{equation}
with $b>a\geq 0$. These integrals can be computed explicitly in terms of elliptic functions. We have\footnote{Our conventions for elliptic functions agree with those of Mathematica.}
\begin{subequations}
\begin{equation}
\begin{aligned}
\label{c1int}
 \CI_1&=\frac{\di}{3}b \left(\left(a^2+b^2\right) {\rm E}_0-\left(a^2+b^2\right) {\rm E}_2
-\left(a^2-b^2\right) \left({\rm F}_2-{\rm K}_0\right)\right)~,
\end{aligned}
\end{equation}
\begin{equation}
\begin{aligned}
 \CJ_1&=\frac{1}{3} \left(a \left(-a^2+b^2\right) {\rm E}_1+a \left(a^2+b^2\right) {\rm K}_1\right)~,
\end{aligned}
\end{equation}
\begin{equation}
\begin{aligned}
 \CI_2&=\frac{\di}{15} b \left(2 \left(a^4-a^2 b^2+b^4\right) {\rm E}_0-2 \left(a^4-a^2 b^2+b^4\right) {\rm E}_2\right.\\&\left.\hspace{2cm}+\left(a^4-3 a^2 b^2+2 b^4\right) \left({\rm F}_2-{\rm K}_0\right)\right)~,
\end{aligned}
\end{equation}
\begin{equation}
\begin{aligned}
 \CJ_2&=\frac{1}{15} a \left(2 \left(a^4+a^2 b^2+b^4\right) {\rm E}_1-\left(2 a^4+3 a^2 b^2+b^4\right) {\rm K}_1\right)~,
\end{aligned}
\end{equation}
\begin{equation}
\begin{aligned}
 \CI_3&=\frac{\di}{35}b \left(2 \left(a^6-5 a^4 b^2-5 a^2 b^4+b^6\right) {\rm E}_0-2 \left(a^6-5 a^4 b^2-5 a^2 b^4+b^6\right) {\rm E}_2\right.\\&\hspace{2cm}\left.+\left(a^6+8 a^4 b^2-11 a^2 b^4+2 b^6\right) \left({\rm F}_2-{\rm K}_0\right)\right)~,
\end{aligned}
\end{equation}
\begin{equation}
\begin{aligned}
 \CJ_3&=\frac{1}{35} \left(2 a \left(-a^6-5 a^4 b^2+5 a^2 b^4+b^6\right) {\rm E}_1+a \left(a^2+b^2\right) \left(2 a^4+9 a^2 b^2-b^4\right){\rm K}_1 \right)~,
\end{aligned}
\end{equation}
\end{subequations}
where we abbreviated complete and incomplete elliptical functions as follows:
\begin{equation}
\begin{aligned}
 {\rm E}_0&:=\text{E}\left(\frac{a^2}{b^2}\right)~,&{\rm E}_1&:=\text{E}\left(-\frac{b^2}{a^2}\right)~,&{\rm E}_2&:=\text{E}\left(\arcsin\left(\frac{b}{a}\right)\Big|\,\frac{a^2}{b^2}\right)~,\\&&&&{\rm F}_2&:=\text{F}\left(\arcsin\left(\frac{b}{a}\right)\Big|\,\frac{a^2}{b^2}\right)~,\\
 {\rm K}_0&:=\text{K}\left(\frac{a^2}{b^2}\right)~,&{\rm K}_1&:=\text{K}\left(-\frac{b^2}{a^2}\right)~.
\end{aligned}
\end{equation}

\subsection{Further integrals}

The integrals given below are used in the analysis of phase III. Recall our approximation for the collective field:
\begin{equation}
 \phi_0(\lambda)=\left\{\begin{array}{ll}\dfrac{1}{\pi}\sqrt{p_0-\alpha(\lambda-\lambda_{\rm min})^2}&\mbox{for }\lambda\in\CI~, \\[0.2cm]
0 & \mbox{else}~.
\end{array}\right.
\end{equation}
The zeros of this collective field define $\CI=[\lambda_L,\lambda_R]$ and are located at
\begin{equation}
 \lambda_L=\lambda_{\rm min}-\sqrt{\frac{p_0}{\alpha}}\eand\lambda_R=\lambda_{\rm min}+\sqrt{\frac{p_0}{\alpha}}~.
\end{equation}
Assuming that the expression under the square root in $\phi_0$ is positive for $\lambda\in[\lambda_L,\lambda_R]$, we have:
\begin{equation}
 \begin{aligned}
    \int_{\lambda_L}^{\lambda_R}\dd \lambda\,\phi_0(\lambda)=\frac{p_0}{2\sqrt{\alpha}}~,~~~\int_{\lambda_L}^{\lambda_R}\dd \lambda\,\phi_0(\lambda)(p_0-\alpha(\lambda-\lambda_{\rm min})^2)^2=\frac{3p_0^2}{8\sqrt{\alpha}}~,\\
    \int_{\lambda_L}^{\lambda_R}\dd \lambda\,\phi_0(\lambda)\lambda=\frac{p_0\lambda_{\rm min}}{2\sqrt{\alpha}}~,~~~
    \int_{\lambda_L}^{\lambda_R}\dd \lambda\,\phi_0(\lambda)\lambda^2=\frac{p_0(p_0+4\alpha\lambda_{\rm min}^2))}{8\alpha^{3/2}}~.~~~
 \end{aligned}
\end{equation}

\pdfbookmark[1]{References}{refs}\label{refs}

\end{document}